\documentclass[aps,twocolumn]{revtex4}

\usepackage{graphicx}
\usepackage{dcolumn}
\usepackage{bm}

\begin{document}
\title{On the nature of the spin-polarized hole states in a quasi-two-dimensional GaMnAs ferromagnetic layer}
\author{E. Dias Cabral}
\affiliation{Institute de F\'\i sica, Universidade do Estado do
Rio de Janeiro, 20.500-013  Rio de Janeiro, R.J., Brazil}
\author{M. A. Boselli}
\affiliation{Departamento de F\'\i sica, Universidade Federal de
Ouro Preto,  35400-000 Ouro
Preto, M.G., Brazil}
\author{A. T. da Cunha Lima}
\affiliation{Universidade Veiga de Almeida,  28905-970 Cabo Frio, RJ, Brazil}
\author{A. Ghazali (in memoriam)}
\affiliation{Institut des NanoSciences de Paris, 75015 Paris, France}
\author{I. C. da Cunha Lima}
\affiliation{Instituto de F\'\i sica, Universidade do Estado do
Rio de Janeiro, 20.500-013 Rio de Janeiro, R.J., Brazil, and\\
Departamento de F\'\i sica, Universidade Federal de Ouro Preto,
 35400-000 Ouro Preto, M.G., Brazil}

\date{\today}

\begin{abstract}
A self-consistent calculation of the density of states and the spectral density
function is performed in a two-dimensional spin-polarized hole system based on a 
multiple-scattering approximation. Using parameters corresponding to GaMnAs thin layers, a wide range of 
Mn concentrations and hole densities have been explored to understand the nature, localized or extended, of the spin-polarized holes at the Fermi level for several values of the average magnetization of the Mn system. We show that, for a certain interval of Mn and hole densities, an increase on the  magnetic order of the Mn ions come together with a change of the nature of the states at the Fermi level. This fact provides a delocalization of spin-polarized extended states anti-aligned to the average Mn magnetization, and a higher spin-polarization of the hole gas. These results are consistent with the occurrence of ferromagnetism with relatively high transition temperatures observed in some thin film samples and multilayered structures of this material.

\end{abstract}


\maketitle

The possibility of having diluted magnetic semiconductor (DMS) nanostructures 
based on GaAs opens a wide range
of potential applications such as integrated magneto-optoelectronic
devices \cite{david}.  In the Ga$_{1-x}$Mn$%
_{x}$As alloy Mn is, in fact, a strong $p$ dopant,
 the free hole concentration reaching even
$10^{20-21}cm^{-3}$. The $3d$ level is half filled with five electrons, in such a way that it
carries a spin $5\hbar /2$. At small Mn concentrations, the alloy is a
paramagnetic insulator. As $x$ increases it becomes ferromagnetic,
going through a non-metal-to-metal transition for
higher concentrations, and keeping its ferromagnetic phase. Above $%
7\%$, the alloy becomes a ferromagnetic insulator.\cite{matsukura}
In the metallic phase, depending on the value of $x$, the
temperature of the ferromagnetic transition is observed in the
range of 30-160 K. The occurrence of ferromagnetism in (Ga,Mn)As thin films  and
(Ga,Mn)As/GaAs superlattices \cite{sado1} made it clear that
a deeper theoretical investigation of such nanostructures is required. Recent calculations \cite{we1,we2}
performed  in GaMnAs/GaAs multilayered structures show the interplay of magnetic order and spin-polarization of free carries occurrying in these systems.

In this work we study the roles of disorder and spin-polarization on determining the nature of the spin-polarized states at the Fermi level in a 2D hole system. We aim to obtain information about the nature of the states of carriers in the metallic and ferromagnetic phase of a thin layer of GaMnAs.  We obtain the density of states  (DOS) by a multiple-scattering approximation in the Klauder approach \cite{klauder} developed by Serre and Ghazali \cite{ghaz1}, and we consider the Zeeman splitting (ZS) equivalent to the separation of the spin aligned and spin anti-aligned subbands resulting of the Kondo-like interaction with the localized magnetic moments. We show that by
analyzing the spectral density function (SDF) we can infer, for a given average magnetization, the nature of the spin-polarized states at the Fermi level. Details  of the multiple-scattering treatment used by Serre and Ghazali can be found in Ref. \onlinecite{ghaz1}.

The scattering potential due to a system of $N_i$ impurities per unit volume
 at sites $\textbf{R}_i$ is given by $
U(\textbf{q})=\rho_{\textrm{imp}}(q)v(\textbf{q})
$, where we used the Fourier transforms of the impurity density
$\rho{_{\textrm{imp}}}(\textbf{r})=\sum_i\delta(\textbf{r}-\textbf{R}_i)
$, and the screened Coulomb potential due to the ionized impurity $v(r)$.  The expansion  implies in performing averages
on products of impurity densities, using the technique by Kohn and
Luttinger \cite{average}. The multiple scattering approximation
consists in selecting from the self-energy insertions  those terms
consisting of the scattering which occur several times by the same
impurity. This is different from the Born approximation used in systems with low concentration of impurities, where only double scattering is considered. The multiple-scattering approach reproduces the correct result even for the very diluted limit.

After performing the configuration average the Green's function   $\overline{G(\textbf{p},E)}$  is diagonal and obeys the Dyson equation with the use of the self-energy $\Sigma(\textbf{p},E)$:
\begin{equation}
\overline{G(\textbf{p},E)}=G^0(\textbf{p},E)+\Sigma(\textbf{p},E)\overline{G(\textbf{p},E)}.
\label{dysonG}
\end{equation}
$ G^0(\textbf{p},E)$ represents the unperturbed Green function.  It is  convenient to define the vertex function by its own Dyson equation in a $d$-dimensional system:
\begin{eqnarray}
K(\textbf{k},\textbf{q};E)=&&\frac{1}{(2\pi)^d}\int
d^d\textbf{q}'v(\textbf{q}'-\textbf{q})\overline{G(\textbf{k}+\textbf{q}')}\times \nonumber \\
&&[N_iv(-\textbf{q}')+K(\textbf{k},\textbf{q}';E)]. \label{presque}
\end{eqnarray}

At this point, in order to make the expansion useful, a change of
variables is performed, $ \textbf{k}+\textbf{q}' \rightarrow \textbf{q}_1 $, and
$\textbf{k}+\textbf{q} \rightarrow  \textbf{q}'_1$.
Next we define the vertex function $K_1(\textbf{k},\textbf{q}_1;E)
\equiv K(\textbf{k},\textbf{q}_1-\textbf{k};E)$, which obeys the
following Dyson equation:
\begin{eqnarray}
K_1(\textbf{k},\textbf{q}_1;E)=&&\frac{1}{(2\pi)^d}\int
d^d\textbf{q}'_1v(\textbf{q}_1-\textbf{q}'_1)\overline{G(\textbf{q}'_1)}\times \nonumber \\
&&[N_iv(\textbf{k}-\textbf{q}'_1)+K_1(\textbf{k},\textbf{q}'_1;E)],
\label{voila}
\end{eqnarray}
In terms of this new vertex function, the impurity self-energy
becomes:
\begin{equation}
\Sigma_{\textrm{ei}}(\textbf{k},E)=K_1(\textbf{k},\textbf{k};E).
\label{sigk1}
\end{equation}
To make this formalism operational, one more change is necessary,
by defining the function $U$:
\begin{equation}
U(\textbf{k},\textbf{q};E)\equiv
K_1(\textbf{k},\textbf{q};E)+N_iv(\textbf{k}-\textbf{q}).
\label{defU}
\end{equation}
In that case, Eq.(\ref{voila}) becomes:
\begin{eqnarray}
U(\textbf{k},\textbf{q};E)=N_iv(\textbf{k}-\textbf{q})+\nonumber \\
\frac{1}{(2\pi)^d}\int
d^d\textbf{q}'_1v(\textbf{q}'-\textbf{q})\overline{G(\textbf{q}')}
U(\textbf{k},\textbf{q}';E), \label{voila1}
\end{eqnarray}
given rise to a linearized matrix equation:
\begin{equation}
[I-\tilde{v}\tilde{G}]\tilde{U}=N_i\tilde{v}. \label{matrixeq}
\end{equation}
The sign tilde is used to identify a matrix. Therefore, the
problem is reduced to a simple linearized matrix equation of the
kind $\tilde{A}.\tilde{X}=\tilde{B}$, where $\tilde{B}=N_i\tilde{v}$
is a fixed matrix, while $\tilde{A}=[I-\tilde{v}\tilde{G}]$, and
$\tilde{X}=\tilde{U}$ changes at each iteration in the
self-consistent process. We start with a free-particle Green's
function and obtain $\tilde{U}$. Next, with Eqs.(\ref{sigk1}) and
(\ref{defU}), we calculate the electron-impurity self-energy, the
Green's function for the next iteration, and so on.

\begin{figure}[t]
 \includegraphics[angle=-90,width=\columnwidth]{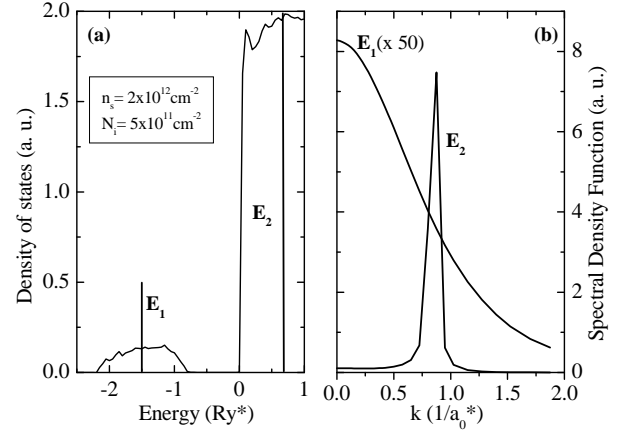}
\caption{DOS(a) for holes in p-doped GaAs as function of energy (1Ry*=55.3 meV), and SDF (b) at energies $E_1=-1.5$ Ry*, $E_2=0.75$ Ry* as function of wavevector (a$^*_0$=21 \AA).}
\label{fig1}
\end{figure}

The self-consistent calculation is performed for a two-dimensional hole gas of
areal density $n_s$ submitted to Coulomb scattering by a negative
ionized impurity system of concentration $N_i$. These two data are
considered as independent parameters. This is important in the present context, since it is known that the density of free carriers in the ferromagnetic GaMnAs samples is just a fraction of the concentration of Mn. The effective mass and the dielectric constant are assumed as  $m_h=0.62$ (heavy hole) and $\kappa_0=12.35$.
At low impurity concentration, independently of the hole
density, an impurity band detaches from the conduction band where
a tail appears. Also, we have
confirmed that at very low impurity concentration and $N_i=n_s$, the
impurity band width tends to zero and becomes centered at
-4Ry$^*$, the well known two-dimensional single impurity $1s$
state. The DOS of a characteristic system with $N_i=2\times 10^{12}$cm$^{-2}$ and 
$n_s=5\times 10^{11}$cm$^{-2}$ is shown in Fig.\ref{fig1}.
Chosing the energies $E_1$ and $E_2$ respectively in the middle of the impurity band and inside the conduction band away from its tail, we  calculated the SDF, $A(\textbf{k},E)=2\Im{\overline{G(\textbf{p},E)}}$, at these energies and we observed that they have typical shapes corresponding to localized and extended states.

\begin{figure}
 \includegraphics[angle=-90,width=\columnwidth]{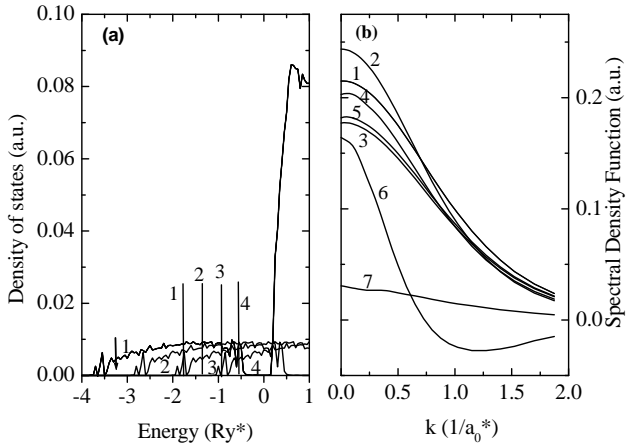}
\caption{Results for $n_s=5\times 10^{11}$cm$^{-2}$ and $N_i=2\times 10^{12}$cm$^{-2}$. In part (a) curve 1 represents the DOS of the anti-aligned  spins,  held fixed. Curves 2, 3, and 4 represent the shifted DOS for aligned spins due to Zeeman splittings 50, 100 and 150 meV. The corresponding Fermi levels are indicated. In part (b), curve 1 corresponds to either aligned or anti-aligned SDF at the Fermi level with zero magnetization, curves 2 and 3 to anti-aligned and aligned spins with a splitting of 50 meV, 4 and 5 idem for 100 meV, and 6 and 7 for 150 meV.}
\label{fig2}
\end{figure}

\begin{figure}
 \includegraphics[angle=-90,width=\columnwidth]{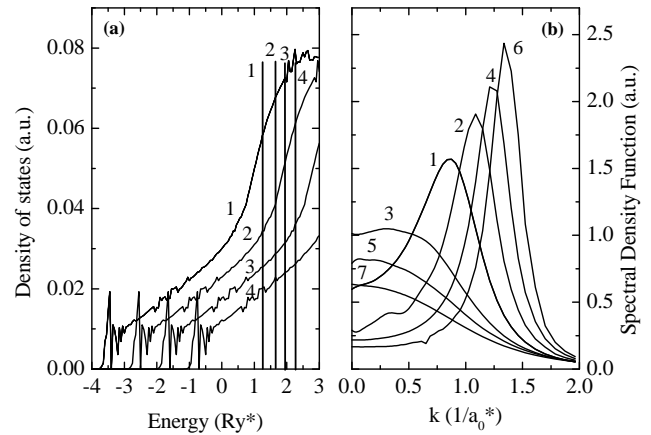}
\caption{Same as above for $n_s=5\times 10^{12}$cm$^{-2}$ and $N_i=1\times 10^{13}$cm$^{-2}$}
\label{fig3}
\end{figure}

Next we take into account the localized magnetic moments at the Mn sites. We treat them assuming a homogeneous magnetization. Then, we obtain an effective magnetic potential \cite{we1} $V_{mag}=-\frac{x}{2}N_0\beta <M>\sigma$, where $\sigma=\pm1$, $x$ is the Mn doping factor, $<M>$ is the average magnetization which for fully aligned moments is  $<M>=5\hbar/2$,  and $N_0\beta$ is the exchange potential for holes, $N_0\beta=-1.2eV$, according to Ref. \onlinecite{matsukura}. If $x=0.05$, $V_{mag}$ introduces a Zeeman splitting of 150 meV for fully magnetic ordered samples. We calculated the Fermi level and the SDF for each spin polarization (aligned and anti-aligned to the average magnetization) corresponding to Zeeman splittings of 50, 100 and 150meV. First we performed the calculation for $n_s=5\times 10^{11}$cm$^{-2}$ and $N_i=2\times 10^{12}$cm$^{-2}$ The results are shown in Fig. \ref{fig2}. The calculation was repeated for 
$n_s=5\times 10^{12}$cm$^{-2}$ and $N_i=1\times 10^{13}$cm$^{-2}$ in  Fig. \ref{fig3}. The DOS corresponding to the anti-aligned spin is fixed and that of the aligned spin is displaced to the right in the energy scale according to the Zeeman splitting. In both cases we see that the Fermi level increases with the
separation. The shape of the spin-dependent SDF does not change considerably as the average magnetization increases in the sample with a lower impurity concentration. However the sample with higher Mn concentration appearing in Fig. \ref{fig3} shows a change in the nature of the states at the Fermi level, even for a splitting of 50 meV. Curve 1 corresponds to zero splitting, where the two SDF coincide. Curve 2 corresponds to the splitting of 50 meV for spin anti-aligned, curve 3 for spin aligned. The former shows an extended character, the latter is localized. As the splitting increases, the SDF for anti-aligned spins becomes  sharper, while for aligned spins it spreads in the k-space, with the maximum approaching $k=0$.

Therefore, if the impurity concentration is below a certain threshold, the increase of the magnetization does not lead to a significant change in the localized character of the spin-polarized states. Above this threshold we reach conditions allowing the occurrence of  extended states at the Fermi level - more properly extended states in a very dirty metal - which are very sensitive to the average magnetization. As the average magnetization increases, a sudden change of the aligned spin states occur, becoming localized, the extended character of the anti-aligned spins becomes more and more pronounced, and the spin polarization of the gas of holes increases.

To conclude, this work approaches the problem of the occurrence of  ferromagnetism in GaMnAs thin films  pointing to the entaglement of the average magnetization with the extended or localized characters of the spin-polarized hole states. We do not try to explain the occurrence of the ferromagnetic order. But, on the other hand, we demonstrate that for a convenient range of Mn concentration the increase of the magnetic order goes together with the increase of the extended character of the majority of the spin-polarized carriers. This is in agreement with results obtained in GaMnAs/GaAs thin quantum wells and multilayered structures \cite{we1,we2} showing that in the metallic phase a strong spin-polarization of the ``free carriers'' in the regions where the Mn impurities are located is responsible for the magnetic order in that samples  with relatively high transition temperatures.

This work was partially supported by CNPq (ESN support and
research grant), FAPEMIG, FAPESP and FAPERJ. ICCL is grateful for
the hospitality of Prof. M. W. Wu group at the USTC, Hefei, Anhui,
China.

\end{document}